# Hybrid electrochromic device with Tungsten oxide ($WO_{3-x}$) and nafion membrane: performance with varying tungsten oxide thickness

K Uday Kumar[1], S D Bhat[2], V V Giridhar[2] and A Subrahmanyam[1]

*1 Semiconductor laboratory, Department of Physics, Indian Institute of Technology Madras, Chennai, 600036, India*

*2 CSIR-Central Electrochemical Research Institute-Madras Unit, CSIR Madras Complex, Chennai 600 113, India*

**Abstract:**

Electrochromic devices, which dynamically change color under the applied potential, are widely studied because of its wide range of applications such as energy-efficient smart windows, rear view mirrors and display devices etc. In this study we are reporting four layer electrochromic device based on tungsten oxide as a electrochromic layer and nafion membrane as a ionic conducting layer. Nafion membranes are generally used in fuel cell applications because of its high ionic conductivity and high optical transparency which is suitable for electrochromic device to attain higher efficiencies. We have prepared an electrochromic device by sandwiching ITO coated glass and $WO_3$ coated ITO thin film between nafion membrane. The overall structure of the device is Glass/ITO/$WO_3$/Nafion/ITO/Glass. We deposited tungsten oxide thin films with different thickness on ITO coated glass substrate at room temperature by using reactive DC Magnetron sputtering and we studied the performance of the electrochromic device with the function of thickness. We have observed that electrochromic efficiency is increasing with increase in the tungsten oxide layer thickness. The efficiency of the device increased from 24.8 $cm^2/C$ to 184.3 $cm^2/C$.

**Key words**: Tungsten oxide, electrochromism, nafion and sputtering.

**Introduction:**

Electrochromic devices, which dynamically changes colour under the applied electric potential, are widely studied because of their emerging applications such as energy-efficient smart windows [1-6], electronic read / write paper[7], rear view mirrors [8], display devices[9] and in medical devices etc. With the emerging concept of electro-chromic smart window, it is anticipated that the transmitted intensity of the day light entering into the window / façade of the building can be controlled to any desirable level with the applied electric potential, there by even the heat also can be reduced. There are several organic and inorganic materials, such as the transition metal oxides, Prussian blue, viologens, the conducting polymers, exhibit this electrochromic phenomenon. Among the several materials explored for the electrochromic applications, tungsten oxide ($WO_{3-x}$) has attracted significant attention.

In the electrochromic material, when a proton ($H^+$) or Lithium ($Li^+$) is introduced into the tungsten oxide lattice (by applying electric potential), the valence state of the tungsten ions change from $W^{6+}$ to $W^{5+}$ or $W^{4+}$, depending upon the available sites. For tungsten oxide, the coloration is explained by the optical transitions[10]:

$$h\nu + W^{5+}(A) + W^{6+}(B) \leftrightarrow W^{6+}(A) + W^{5+}(B) \qquad (1)$$

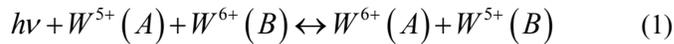

where A and B indicates two neighbouring W sites, and represents the photon energy. Details of these transitions are explained by well-known Polaron model [Ref]. When the voltage is reversed, these electrons and ions come out of the tungsten oxide lattice and as a result the tungsten oxide lattice valance state returns to its normal position (bleached state). The optical transmission, the density of intercalated charge and the distance through which the proton / lithium ion traverses in the tungsten oxide lattice depends upon the thickness of the electrochromic (tungsten oxide) layer. An optimum thickness is desired for an efficient device. There are very few reports available on the effect of thickness of tungsten oxide layer [11, 12].

The inorganic solid-state electrochromic device consists of four layers: transparent conductor (ITO)/ electro-chromic layer ($WO_3$) / ionic conductor ($Ta_2O_5$)/ transparent conductor (ITO) [11]. A simple polymeric electro-chromic device consists of: conductive polymer / polymer electrolyte (PEDOT) / conductive polymer [13]. There are several combinations of the inorganic – organic (hybrid) electro-chromic device. In the present study, the device configuration used is: ITO/ $WO_3$/ Nafion membrane /ITO.

The coloration efficiency, CE, (at a given wavelength) is expressed as the ratio between the optical contrast in the bleached state ($T$b) and the colored state ($T$c) and the charge density ($Q/A$) intercalated [14]

$$CE = \frac{1}{\left(Q/A\right)} \ln\left(\frac{T_b}{T_c}\right)$$

Interestingly, for a specific intercalated charge density, the Tb and Tc have a spectral dependence, thus, CE can have different values at different wavelengths. For a realistic comparison, one should choose a specific wavelength; say 550 nm, of the visible optical spectrum.

Switching time (t) is defined as: the time required for the coloring bleaching process of an electrochromic device. Coloration time (tc) is defined as the time required to switch from the bleached state to the colored state.

In the present study, we report the fabrication and electrochromic properties of a hybrid electrochromic device consisting of tungsten oxide ($WO_{3-x}$) as electrochromic layer and the commercially available nafion 117 membrane (183 µm thick) is the ion conducting layer. The effect of the tungsten oxide thickness (40 nm to 200 nm) on the electro-chromic performance of the hybrid device is studied.

**Experimental methods:**

The schematic of the hybrid device is shown in Fig. 1. The electrochromic device consists of: ITO (400 nm)/ Nafion (183 µm) / $WO_{3-x}$ (44 nm to 200 nm)/ ITO (400 nm). The device is prepared on cleaned soda lime glass (Fischer Scientific).

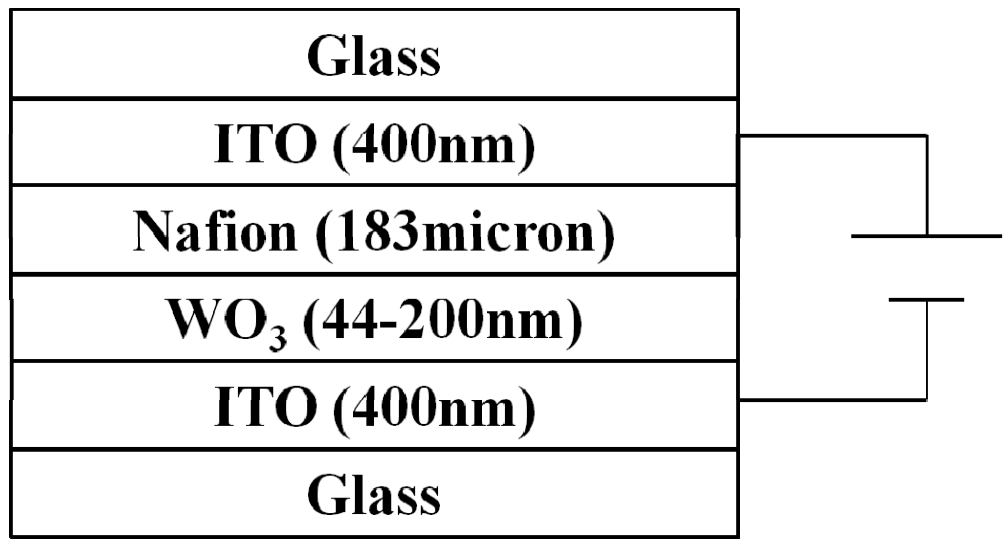

Fig 1: Schematic cross section of electrochromic device based on nafion membrane.

The ITO (400 nm) thin films are deposited at room temperature (300 K) on cleaned glass substrates by conventional reactive DC magnetron sputtering with metallic target of composition In:Sn 90:10; the size of the target is 38mm x 12.7mm. The growth chamber is initially evacuated to $5 \times 10^{-5}$ mbar and the pure argon and pure oxygen gases are introduced into the growth chamber through mass flow controllers at flow rates of 24 sccm and 08 sccm respectively; the deposition pressure in the chamber is 3 x $10^{-3}$ mbar. The In:Sn target is powered with arc controlled Advanced Energy power supply (MDX 1KW) at 0.55 W/cm$^2$. The sheet resistance of ITO is 20 Ω per square and has the optical transmission ~ 89% (at 550 nm).

The $WO_{3-x}$ thin films have been deposited on the ITO thin films by reactive DC magnetron sputtering with pure metallic tungsten target (38mm x 12.7mm). The deposition pressures remain the same at 3.0 x $10^{-3}$ mbar with flow rates of argon and oxygen at 20 sccm and 10 sccm respectively. The magnetron power to the tungsten target is 0.55 W/cm$^2$. The films are coated for different deposition times: 30, 45, 60, 90 and 120 minutes to obtain different thicknesses, accordingly, the samples are labelled: 30-G, 45-G, 60-G, 90-G and 120-G respectively

The thickness of the ITO and $WO_{3-x}$ thin films is measured by profilometry (Bruker 3D Non-contact Profiler).

The nafion 117 membrane (procured commercially) is placed on $WO_{3-x}$ layer by using nafion solution glue to make electro-mechanical contact between $WO_{3-x}$ and nafion; this electro-mechanical contact is made by pressing the two sandwiched layers with a mechanical press. Finally another ITO (400 nm) coated glass is placed on this nafion membrane. The overall structure of the device is Glass/ITO/$WO_{3-x}$/nafion membrane /ITO / Glass.

The ion conducting layer in the present hybrid device is the nafion membrane. The desirable properties of the ion conducting layer are: should have very high ion conductivity and should be optically transparent and if it is a membrane, should have enough mechanical strength to withstand the mechanical pressing. Nafion membrane is a highly proton conductive polymer ionomer, which is generally used in the fuel cell applications [15]. Nafion membrane exhibits high optical transparency and ionic conductivity (0.16 S/cm)[16] and its chemical structure consists of tetrafluoroethylene (TFE) backbone, which is responsible for both the mechanical and the chemical stability, and perfluoroether side chains end-capped with sulfonic acid groups are responsible for its exceptional proton conduction properties [17]. The thickness of the nafion membrane is 183 microns.

The tungsten oxide ($WO_{3-x}$) thin films have been characterized by the conventional techniques: X ray diffraction (Xpert Pro diffractometer with Cu $K_\alpha$ radiation), Raman spectroscopy (Horiba Jobin Yvon Lab Ram with excitation wavelength 532 nm) and optical absorption (Cary 60) in the wavelength range 200 nm to 1100 nm). The hybrid electrochromic device is characterized by the cyclic voltammetry and the chrono amperometry (Biologic SP-300). The coloration efficiency and the cyclability tests also have been conducted.

All the results are reproducible within the experimental error.

**Note on the desirable properties of the electro-chromic device and the electro-chromic layer:**

The desired electrochromic device requirements are: (i) high coloration efficiency, (ii) fast response time for colouring and bleaching and (iii) a near perfect bleached state (no residual charge in the lattice) even after significant number of cycles operation. From the electro-chromic material point of view, and with specific reference to the inorganic metal oxide layers, for an efficient intercalation of charge and for high coloration efficiency, the metal (tungsten) oxide lattice should (i) be an amorphous or porous structure for easy charge transfer, (ii) have a high relative density for high colour contrast and (iii) have the intercalated charge not to leave any residue in the metal oxide lattice [1].

**Results and discussion:**

The thickness of ITO thin films are 400 ± 5 nm.

The X ray diffraction of $WO_{3-x}$ thin films did not show any diffraction peaks; the films are amorphous (figure not given). Amorphous nature of WO3 is desirable for electro-chromic applications.

Fig 2 shows the Raman spectra of tungsten oxide thin films with different deposition times. The most prominent peak at 790 cm$^{-1}$ is attributed to O–W$^{6+}$-O; a broad peak at 950 cm$^{-1}$, the stretching mode of terminal oxygen atoms because of W$^{6+}$ = O, peak corresponding to 267 cm$^{-1}$ can be attributed to the bending vibration of δ(W–O–W) and the peak arising is 478 cm$^{-1}$ is due to the external visible light during the experiment.

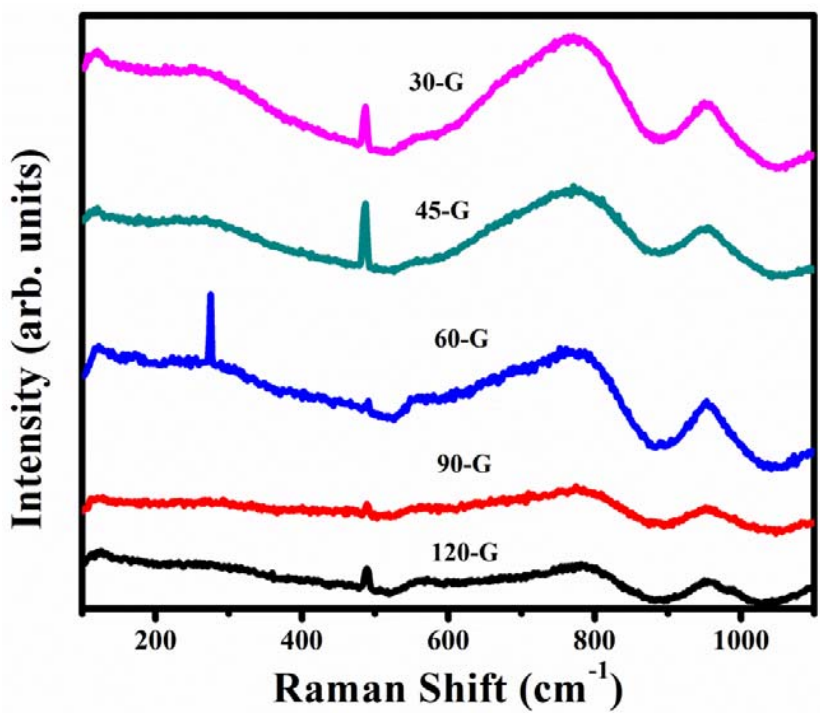

Fig. 2 shows the Raman spectra of the $WO_{3-x}$ thin films.

A transmission spectrum recorded in the wavelength range from 200 nm to 1100 (Fig. 3a). The as deposited pure $WO_{3-x}$ films having high transmittance value range from 75%- 85% in the visible range of the optical spectrum. The reduction of the intensity in the transmission spectra at a wavelength ~370 nm is due the fundamental absorption edge [18]. The transmission spectra of all the films with different thickness are similar in nature. The optical band gap values are calculated from Tau'c s plot (Fig. 3b), are found to be 3.87 eV, 3.71 eV, 3.68 eV, 3.57 eV and 3.56 eV for the films 30-G, 45-G, 60-G, 90-G and 120-G respectively (Table 1). The optical band gap values are higher compared to the reported values [19], can be due to more oxygen deficiency in magnetron sputtered $WO_{3-x}$ thin films. It is also observed that the band gap is increasing with the increasing of film thickness which is due to an increase in the number of carriers (electrons) in the conduction band. The refractive index of the tungsten oxide thin films are measured at 550 nm wavelength observed to be decreasing with the increase of thickness. the decrease in the refractive index is mainly due to the lower density.

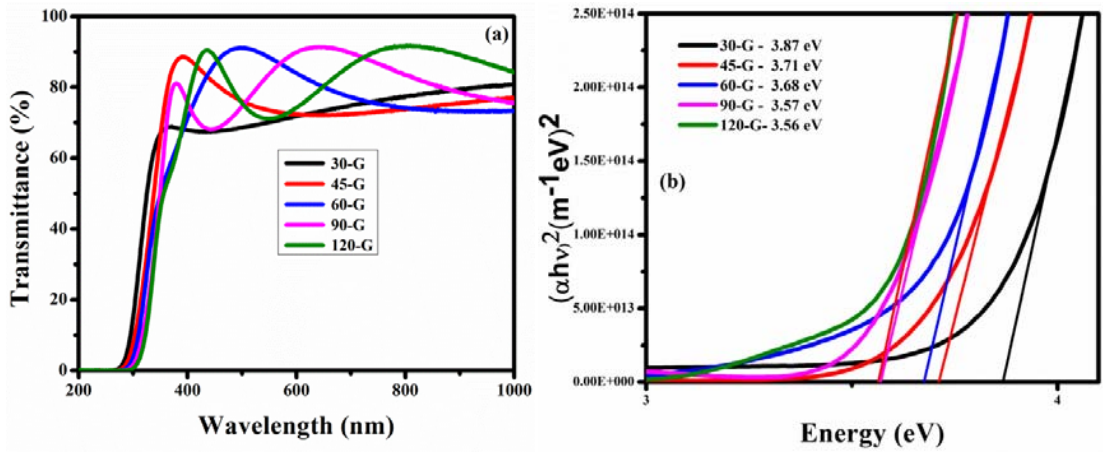

Fig.3 shows (a) the Optical transmission spectra (b) $(\alpha h\nu)^2$ versus $h\nu$ plots for $WO_{3-x}$ thin films for different thicknesses.

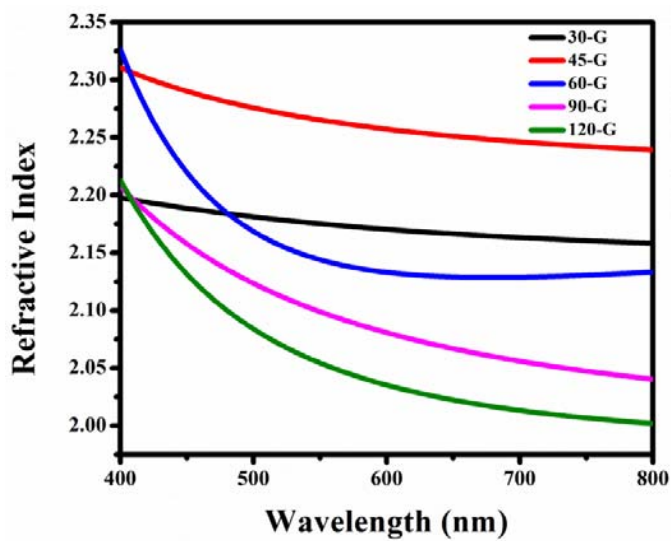

Fig 4: Real part of refractive index as a function of wavelength for different thickness of WO$_{3-x}$ thin films.

**Electrochemical properties of nafion based device:**

Cyclic voltammetry measurements have been carried out on the hybrid electrochromic device (Fig. 5). The active area of the electrochromic devices are 3cm$^2$. The voltage is scanned between -5V to +5V with a scan rate of 100mV/sec. In this device WO$_{3-x}$ layer act as active electrochromic layer and nafion membrane is the ion conducting layer; the top contact ITO act as a counter electrode. During the negative bias, the tungsten oxide layer changes its optical properties from the transparent to the coloured state due to the insertion of protons into the tungsten lattice. These protons supplied by the nafion membrane, diffuse into the tungsten layer because of its acidity. No electrochemical reaction takes place at the ITO electrode side. At 1.2V, the peak observed in all the cyclic voltammograms may be attributed to the optical transitions in the tungsten valence state from W$^{6+}$ state to W$^{5+}$ state. On the other hand, during the bleaching process, when we apply the positive bias voltages to the tungsten oxide layer, the protons will de-intercalate from the tungsten lattice and these protons will move towards the nafion membrane layer. This process takes place till the tungsten oxide become completely bleached. Finally these protons will reach the ITO, now this ITO acts as the counter electrode.

The charge intercalated and de-intercalated into the tungsten lattice is calculated from the cyclic voltammetry curves by integrating the current and time. The charge capacity values for the electrochromic devices are given in the Table 1. Chronoamperometry measurements performed on these hybrid electrochromic devices (Fig. 6) by applying ±5V voltages for 100 sec. The current density values are observed to be increasing with increasing of the thickness of $WO_{3-x}$ layer. The time taken for the conversion of the coloured state to the bleached states have been calculated from these chronoamperometry measurements. The response times for the electrochromic devices are listed in the Table 1. In all the electrochromic devices, the bleaching times are observed to be higher than that of the coloring times.

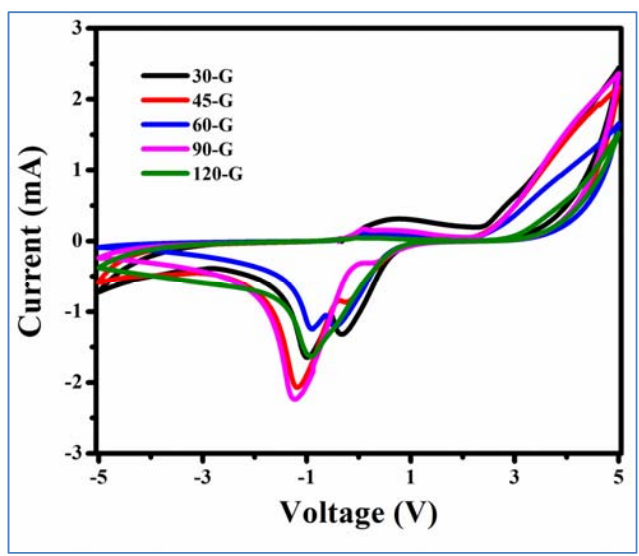

Fig 5: Cyclic voltammograms of nafion membrane based electrochromic devices.

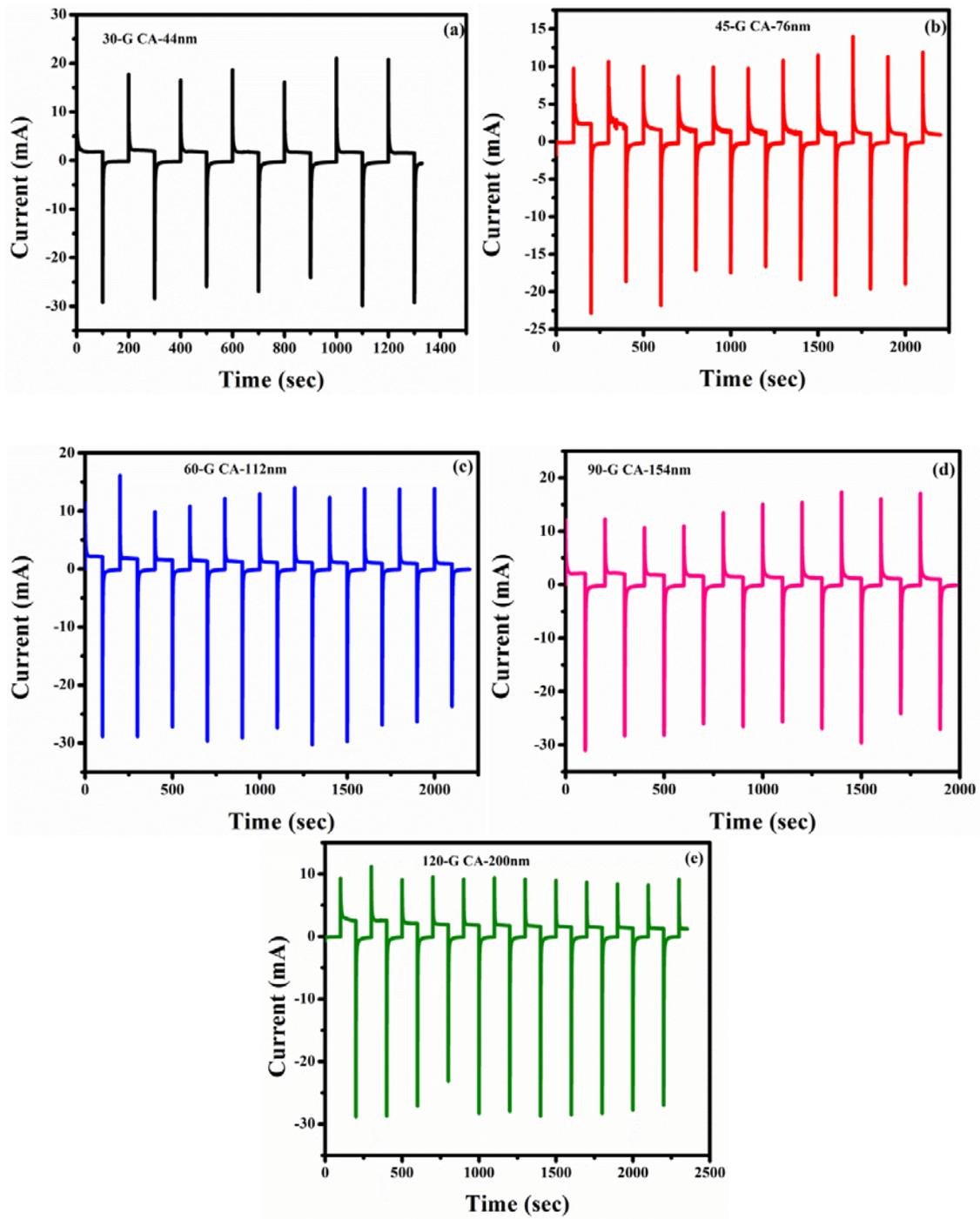

Fig 6: The current versus time responses for nafion membrane based electrochromic devices

The change in the optical transmission of the coloured and the bleached states are measured by the optical spectrometry and the coloration efficiency (CE) values are evaluated at 550nm wavelength (Fig 7). The vales are found to be 24.2%,24.8cm$^2$/C; 29.1%,45.4cm$^2$/C; 31.0%,72.6cm$^2$/C; 41.6%,77.9cm$^2$/C and 37.8%,184.3cm$^2$/C for Devices 30-G, 45-G, 60-G, 90-G and 120-G respectively. Fig 8 shows that, the coloration efficiency values are found to be increasing with increasing thickness of the tungsten oxide layer. Fig 9 shows the coloured and bleached states of the electrochromic device.

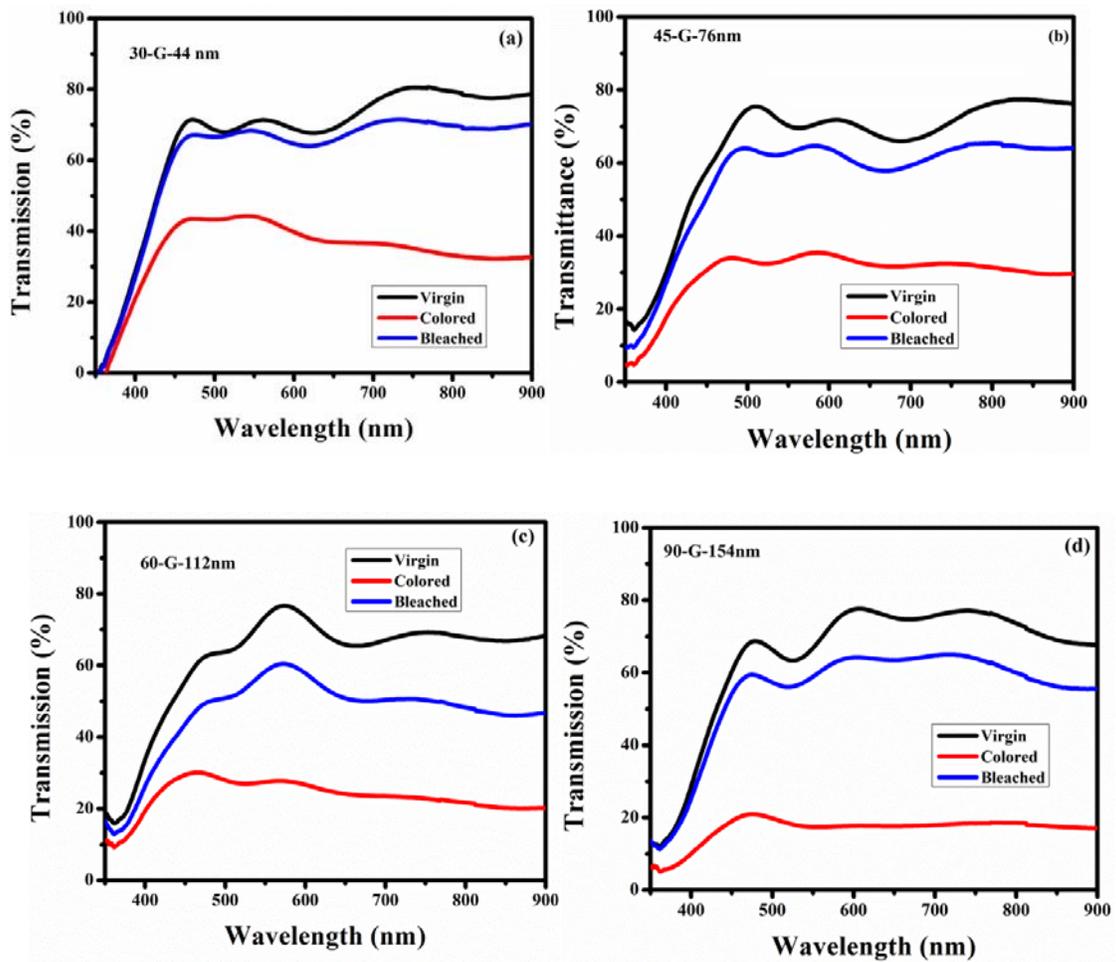

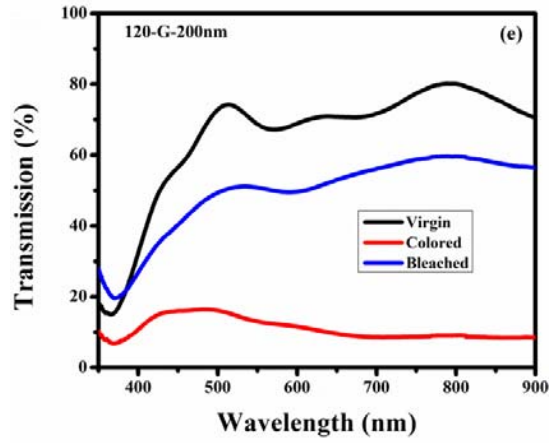

Fig 7: The optical transmission spectra of colored and bleached states of nafion membrane based electrochromic devices.

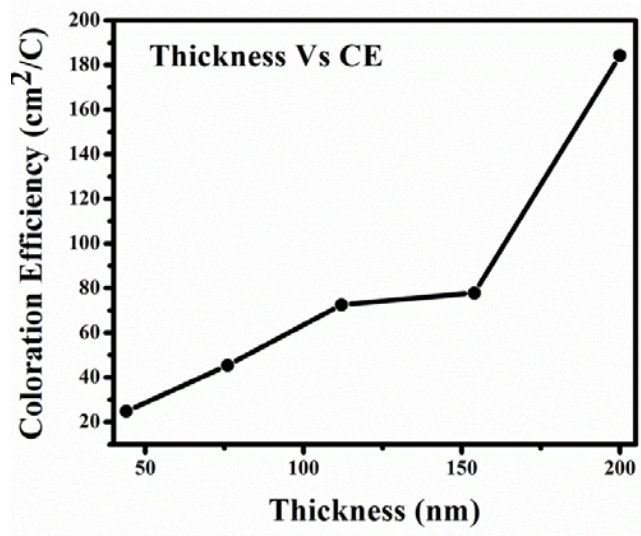

Fig 8: The coloration efficiency of a device as a function of thickness of $WO_{3-x}$ thin films

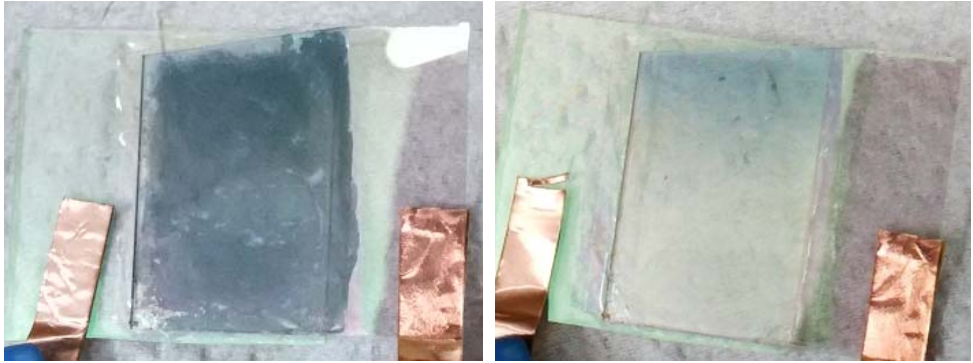

Fig 9: Colored and bleached states of the electrochromic device.

Table 1: Summary of WO$_{3-x}$ layer thickness, transmission change, refractive index, relative density, inserted charge, coloring time, bleaching time and coloration efficiency of nafion membrane based electrochromic devices prepared on glass substrates.

| Sample | WO$_{3-x}$ thickness (nm) | Refractive index @632nm | Relative density @632nm | ΔT @550 nm | Inserted Charge (mC) | Coloring Time (sec) | Coloration Efficiency (cm$^2$/C) | Diffusion coeff (10$^{-12}$cm$^2$/C) |
|---|---|---|---|---|---|---|---|---|
| 30-G | 44±5 | 2.17 | 0.87 | 24.2 | 53.2 | 45 | 24.8 | 0.36 |
| 45-G | 76±5 | 2.25 | 0.90 | 29.1 | 41.0 | 68 | 45.4 | 0.68 |
| 60-G | 112±5 | 2.13 | 0.85 | 31.0 | 31.4 | 62 | 72.6 | 0.21 |
| 90-G | 154±5 | 2.07 | 0.82 | 41.6 | 46.6 | 65 | 77.9 | 0.63 |
| 120-G | 200±5 | 2.03 | 0.80 | 37.8 | 22.3 | 13 | 184.3 | 0.37 |

**Conclusions:** The tungsten oxide thin films have been prepared by reactive DC reactive magnetron sputtering with the varying thickness. The hybrid electrochromic device has been prepared by using nafion membrane. The highest coloration efficiency observed for the film 120-G is about 184.3 cm$^2$/C. The coloration efficiency values were observed to be increasing with the increase of the thickness.

**References**:

[1] K. Bange, Colouration of tungsten oxide films: A model for optically active coatings, Solar Energy Materials and Solar Cells, 58 (1999) 1-131.
[2] Y. Wang, E.L. Runnerstrom, D.J. Milliron, Switchable Materials for Smart Windows, Annual Review of Chemical and Biomolecular Engineering, 7 (2016) 283-304.
[3] C.G. Granqvist, Electrochromics for smart windows: Oxide-based thin films and devices, Thin Solid Films, 564 (2014) 1-38.
[4] J.S.E.M. Svensson, C.G. Granqvist, Electrochromic coatings for "smart windows", Solar Energy Materials, 12 (1985) 391-402.
[5] C.M. Lampert, Electrochromic materials and devices for energy efficient windows, Solar Energy Materials, 11 (1984) 1-27.
[6] C.G. Granqvist, Handbook of inorganic electrochromic materials, Elsevier, 1995.
[7] D. Corr, U. Bach, D. Fay, M. Kinsella, C. McAtamney, F. O'Reilly, S.N. Rao, N. Stobie, Coloured electrochromic "paper-quality" displays based on modified mesoporous electrodes, Solid State Ionics, 165 (2003) 315-321.
[8] F.G.K. Baucke, Electrochromic mirrors with variable reflectance, Solar Energy Materials, 16 (1987) 67-77.
[9] T. Oi, Electrochromic Materials, Annual Review of Materials Science, 16 (1986) 185-201.
[10] D. Dini, F. Decker, E. Masetti, A comparison of the electrochromic properties of WO3 films intercalated with H+, Li+ and Na+, Journal of Applied Electrochemistry, 26 (1996) 647-653.
[11] P.S. Patil, P.R. Patil, S.S. Kamble, S.H. Pawar, Thickness-dependent electrochromic properties of solution thermolyzed tungsten oxide thin films, Solar Energy Materials and Solar Cells, 60 (2000) 143-153.
[12] M.H. Kim, H.W. Choi, K.H. Kim, Thickness Dependence of WO3-x Thin Films for Electrochromic Device Application, Molecular Crystals and Liquid Crystals, 598 (2014) 54-61.
[13] D. Mecerreyes, R. Marcilla, E. Ochoteco, H. Grande, J.A. Pomposo, R. Vergaz, J.M. Sánchez Pena, A simplified all-polymer flexible electrochromic device, Electrochimica Acta, 49 (2004) 3555-3559.
[14] K.U. Kumar, S.M. Dhanya, A. Subrahmanyam, Flexible electrochromics: magnetron sputtered tungsten oxide (WO 3– x ) thin films on Lexan (optically transparent polycarbonate) substrates, Journal of Physics D: Applied Physics, 48 (2015) 255101.
[15] O. Diat, G. Gebel, Fuel cells: Proton channels, Nat Mater, 7 (2008) 13-14.
[16] S. Slade, S.A. Campbell, T.R. Ralph, F.C. Walsh, Ionic Conductivity of an Extruded Nafion 1100 EW Series of Membranes, Journal of The Electrochemical Society, 149 (2002) A1556-A1564.
[17] H. Kamal, A.A. Akl, K. Abdel-Hady, Influence of proton insertion on the conductivity, structural and optical properties of amorphous and crystalline electrochromic WO3 films, Physica B: Condensed Matter, 349 (2004) 192-205.
[18] K. Gesheva, A. Szekeres, T. Ivanova, Optical properties of chemical vapor deposited thin films of molybdenum and tungsten based metal oxides, Solar Energy Materials and Solar Cells, 76 (2003) 563-576.

[19] H. Kaneko, F. Nagao, K. Miyake, Preparation and properties of the dc reactively sputtered tungsten oxide films, Journal of Applied Physics, 63 (1988) 510-517.